\RequirePackage[T1]{fontenc}
\documentclass[11pt]{article}
\usepackage[utf8]{inputenc}
\usepackage{typearea}\typearea{13}

\usepackage{amsmath,amsfonts,amssymb,upgreek,mathtools,epsfig}
\usepackage{cite}
\usepackage[svgnames]{xcolor}
\usepackage[colorlinks,citecolor=DarkGreen,linkcolor=FireBrick,urlcolor=FireBrick,linktocpage,breaklinks=true]{hyperref}

\makeatletter
\long\def\@makecaption#1#2{{\small
\advance\leftskip1cm
\advance\rightskip1cm
\vskip\abovecaptionskip
\sbox\@tempboxa{#1: #2}%
\ifdim \wd\@tempboxa >\hsize
 #1: #2\par
\else
\global \@minipagefalse
\hb@xt@\hsize{\hfil\box\@tempboxa\hfil}%
\fi
\vskip\belowcaptionskip}}
\makeatother

\makeatletter
\def\renewtheorem#1{%
  \expandafter\let\csname#1\endcsname\relax
  \expandafter\let\csname c@#1\endcsname\relax
  \gdef\renewtheorem@envname{#1}
  \renewtheorem@secpar
}
\def\renewtheorem@secpar{\@ifnextchar[{\renewtheorem@numberedlike}{\renewtheorem@nonumberedlike}}
\def\renewtheorem@numberedlike[#1]#2{\newtheorem{\renewtheorem@envname}[#1]{#2}}
\def\renewtheorem@nonumberedlike#1{  
\def\renewtheorem@caption{#1}
\edef\renewtheorem@nowithin{\noexpand\newtheorem{\renewtheorem@envname}{\renewtheorem@caption}}
\renewtheorem@thirdpar
}
\def\renewtheorem@thirdpar{\@ifnextchar[{\renewtheorem@within}{\renewtheorem@nowithin}}
\def\renewtheorem@within[#1]{\renewtheorem@nowithin[#1]}
\makeatother

\newcommand{\nl}{\notag\\}
\def\eq#1\en{\begin{equation}#1\end{equation}}  
\def\eqsplit#1\ensplit{
	\begin{equation}\begin{split}#1\end{split}\end{equation}
	}
\def\eqalign#1\enalign{
	\begin{align}#1\end{align}
	}
\def\eqa#1\ena{
	\begin{align}#1\end{align}
	}
\def\eqg#1\eng{
	\begin{gather}#1\end{gather}
}
\def\eqmul#1\enmul{
	\begin{multline}#1\end{multline}
	}
\newcommand{\lb}[1]{\label{e:#1}}
\newcommand{\rlb}[1]{\eqref{e:#1}}     
\newtheorem{theorem}{Theorem}
\newtheorem{T}[theorem]{Theorem}
\renewtheorem{C}[theorem]{Corollary}

\newtheorem{Le}[theorem]{Lemma}
\newtheorem{D}[theorem]{Definition}

\newcommand{\qedm}{\rule{1.5mm}{3mm}}

\newcommand{\snorm}[1]{\Vert#1\Vert}

\newcommand{\sbkt}[1]{\langle#1\rangle}

\newcommand{\bra}[1]{\langle#1|}
\newcommand{\ket}[1]{|#1\rangle}

\newcommand{\sumtwo}[2]%
{\mathop{\sum_{#1}}_{#2}}
\newcommand{\sumthree}[3]%
{\mathop{\mathop{\sum_{#1}}_{#2}}_{#3}}
\newcommand{\sumfour}[4]%
{\mathop{\mathop{\mathop{\sum_{#1}}_{#2}}_{#3}}_{#4}} 
\newcommand{\prodtwo}[2]%
{\mathop{\prod_{#1}}_{#2}}
\newcommand{\mintwo}[2]%
{\mathop{\min_{#1}}_{#2}}
\newcommand{\maxtwo}[2]%
{\mathop{\max_{#1}}_{#2}}
\newcommand{\maxthree}[3]%
{\mathop{\mathop{\max_{#1}}_{#2}}_{#3}}
\newcommand{\limtwo}[2]%
{\mathop{\lim_{#1}}_{#2}}
\newcommand{\suptwo}[2]%
{\mathop{\sup_{#1}}_{#2}}
\newcommand{\supthree}[3]%
{\mathop{\mathop{\sup_{#1}}_{#2}}_{#3}}
\newcommand{\supfour}[4]%
{\mathop{\mathop{\mathop{\sup_{#1}}_{#2}}_{#3}}_{#4}} 
\newcommand{\inftwo}[2]%
{\mathop{\inf_{#1}}_{#2}}
\newcommand{\infthree}[3]%
{\mathop{\mathop{\inf_{#1}}_{#2}}_{#3}}
\newcommand{\inffour}[4]%
{\mathop{\mathop{\mathop{\inf_{#1}}_{#2}}_{#3}}_{#4}} 

\newcommand\calB{{\cal B}}

\newcommand\calH{{\cal H}}

\newcommand\calK{{\cal K}}

\newcommand\calN{{\cal N}}

\newcommand\calR{{\cal R}}

\newcommand{\bbC}{\mathbb{C}}

\newcommand{\bbN}{\mathbb{N}}

\newcommand{\bbZ}{\mathbb{Z}}

\newcommand{\La}{\Lambda}
\newcommand{\OA}{\mathfrak{A}}
\newcommand{\AL}{\mathfrak{A}^\rL}
\newcommand{\AR}{\mathfrak{A}^\rR}
\newcommand{\AD}{\mathfrak{A}^\rD}
\newcommand{\Az}{\OA_{\{0\}}}
\newcommand{\hilb}{\mathfrak{h}}
\newcommand{\rR}{\mathrm{R}}
\newcommand{\rL}{\mathrm{L}}
\newcommand{\rD}{\mathrm{D}}
\newcommand{\hXi}{\tilde{\Xi}}
\newcommand{\sL}{\sigma^\rL}
\newcommand{\sR}{\sigma^\rR}
\newcommand{\sD}{\sigma^\rD}
\newcommand{\tcK}{\widetilde{\calK}}
\newcommand{\Id}{\mathbb{I}}
\newcommand{\tV}{\tilde{V}}

\newcommand{\ZZ}{\bbZ_2\times\bbZ_2}

\newcommand{\sw}{\sigma\mathrm{\hbox{-}weak}}

\newcommand{\Uo}{{\rm U}(1)}
\newcommand{\ua}{\mathfrak{p}}
\newcommand{\Uu}{\Uo_{\ua}}
\newcommand{\SC}{H^2(G,\Uu)}
\newcommand{\tcR}{\tilde{\calR}}

\numberwithin{equation}{section}

\begin{document}
\begin{flushright}
\footnotesize Draft, April 23, 2020 \\
IPMU-20-0033
\end{flushright}

\noindent
{\Large\bf 
General Lieb-Schultz-Mattis type theorems for quantum spin chains 
}
\par\bigskip

\renewcommand{\thefootnote}{\fnsymbol{footnote}}
\noindent
Yoshiko Ogata\footnote{%
Graduate School of Mathematical Sciences
The University of Tokyo, Komaba, Tokyo 153-8914, Japan.
Supported in part by
the Grants-in-Aid for
Scientific Research, JSPS.},
Yuji Tachikawa\footnote{%
Kavli Institute for Physics and Mathematics of the Universe(WPI), the University of Tokyo,5-1-5 Kashiwanoha, Kashiwa, 275-8583 Japan.
}
and Hal Tasaki\footnote{%
Department of Physics, Gakushuin University, Mejiro, Toshima-ku, 
Tokyo 171-8588, Japan.
}
\renewcommand{\thefootnote}{\arabic{footnote}}\setcounter{footnote}{0}

\begin{quote}
\small
We develop a general operator algebraic method which focuses on projective representations of symmetry group for proving Lieb-Schultz-Mattis type theorems, i.e., no-go theorems that rule out the existence of a unique gapped ground state (or, more generally, a pure split state), for quantum spin chains with on-site symmetry.
We first prove a theorem for translation invariant spin chains that unifies and extends two theorems proved by two of the authors in \cite{OT1}.
We then prove a Lieb-Schultz-Mattis type theorem for spin chains that are invariant under the reflection about the origin and not necessarily translation invariant.
\end{quote}

\tableofcontents

\section{Introduction}
\label{s:intro}
Quantum spin systems have been active topics of research both in theoretical and mathematical physics, see e.g.~\cite{BR1,BR2,Sutherland,ZengChenZhouWenBOOK,TasakiBook}.
The Lieb-Schultz-Mattis theorem \cite[Appendix B]{LSM} and its extensions  e.g.~in \cite{AL,AizenmanNachtergaele1994,OYA,YOA,O,H1,H2,NS,TasakiLSM,BachmannBolsDeRoecFraas2019} are attracting renewed interest partly because of their close relations to topological phases of matter, see e.g.~\cite{ZengChenZhouWenBOOK,TasakiBook}.
More precisely, these theorems state that certain classes of quantum many-body systems with U(1) invariance cannot have a unique ground state accompanied by a nonzero energy gap, while the classification of unique gapped ground states is a central issue in topological condensed matter physics.

The original theorem and early extensions  were based on explicit construction of low-lying excited states above the ground state \cite{LSM,AL,OYA,YOA,TasakiLSM}.
In \cite{O}, where the extension of the theorem to two or higher dimensions was first discussed, Oshikawa directly examined a necessary condition for the existence of a unique gapped ground state.
This rephrasing of the Lieb-Schultz-Mattis theorem was essential for the later development, including the present work.

Recently, in the context of topological condensed matter physics, it was argued that Lieb-Schultz-Mattis type no-go theorems should be valid for quantum many-body systems that only possess certain discrete symmetry \cite{ChenGuWEn2011,PTAV,WPVZ,PWJZ,Watanabe2018}.
In particular it was conjectured by Chen, Gu, and Wen \cite[V.B.4 and  V.C]{ChenGuWEn2011}, as a part of their general classification, that a translation invariant quantum spin chain where the representation of the symmetry  on each site is genuinely projective cannot have a unique gapped ground state.
The statement for chains with time-reversal symmetry was proved by Watanabe, Po, Vishwanath, and Zaletel \cite{WPVZ} within the framework of matrix product states (MPS).\footnote{%
See also \cite{Prakash} and \cite[section~8.3.5]{TasakiBook} for general proofs for MPS.}
In \cite{OT1}, Ogata and Tasaki confirmed the conjecture with full mathematical rigor for general translation invariant quantum spin chains with $\ZZ$ or time-reversal symmetry.
The proof was an extension of the early work of Matsui \cite{Matsui1}, where the method based on the Cuntz algebra was developed.

 \medskip
 
In the present work we essentially complete the study of Lieb-Schultz-Mattis type theorems for quantum bosonic spin chains with discrete on-site symmetry by proving two general theorems.
We first provide a general unified proof for the above mentioned conjecture by  Chen, Gu, and Wen \cite{ChenGuWEn2011}.
See Corollary~\ref{C:trans} and Theorem~\ref{T:trans} below.
We then state and prove a  Lieb-Schultz-Mattis type theorem for spin chains with reflection symmetry.
See Corollary~\ref{C:ref} and Theorem~\ref{T:ref}.
More precisely, we prove that a class of quantum spin chains with certain on-site symmetry and invariance under the reflection about the origin cannot have a unique gapped ground state (or, more generally, a pure split state) when the spin at the origin is half-odd-integral (or, more generally, has a degree-2 cohomology class that is not written in the form $2c$).
This statement previously appeared as a conjecture in a paper by Po, Watanabe, Jian, and Zaletel \cite{PWJZ}.

The proofs of the theorems are closely related to standard ideas in topological condensed matter physics, and does not make use of the Cuntz algebra.
They are based, in an essential manner, on the fact that a unique gapped ground state satisfies the split property, as proven by Matsui \cite{Matsui2}, and that there are projective representations associated to states satisfying the split property, as was noted e.g.~in \cite{Matsui1}.
In \cite{OgataZ2} Ogata showed that the second cohomology class associated to the projective representation is actually an invariant of symmetry protected topological (SPT) phases, in the sense that it is stable under the smooth path of symmetric gapped Hamiltonians, which coincides with the topological index investigated intensively in the context of SPT phases in MPS \cite{Perez-Garcia2008,PollmannTurnerBergOshikawa2010,Schuch,ChenGuWEn2011,ZengChenZhouWenBOOK,TasakiBook}.
We prove the two theorems in a unified manner by using some basic properties of the indices.
As we shall see in section~\ref{s:outlie}, the proofs are straightforward and natural, once some key properties of the indices are given.
The simplicity of the argument suggests that the machinery developed here is the correct language for discussing Lieb-Schultz-Mattis type theorems for quantum spin chains with discrete on-site symmetry.\footnote{%
Similar machinery can also be used to classify unique gapped ground states of a general quantum spin chain with on-site symmetry \cite{OT2}.
}

\medskip

Before proceeding, we pause here to mention that in \cite{OT1} two of the authors of the present paper argued that there is an essential difference between the early Lieb-Schultz-Mattis-type theorems based on the U(1) symmetry and the recent theorems that make use of the projectivity of the representation of the symmetry.
However the two types of theorems may be understood in a unified manner from the view point of quantum anomaly presented in \cite{CHR}.
See the end of  Appendix~\ref{app:Lie} for more details.

%
%
%
%

\paragraph{Two classes of examples}
It may be useful to present two concrete cases of our general theorems in the context of standard quantum spin chains with $\ZZ$ or time-reversal symmetry.

We consider a quantum spin system on the infinite chain $\bbZ$.
Let $\bar{S}_x\in\{\frac{1}{2},1,\frac{3}{2},\ldots,\bar{S}_{\rm max}\}$ be the spin quantum number associated with site $x\in\bbZ$, where $\bar{S}_{\rm max}\in\bbN/2$ is an arbitrary constant.
The system is described by the formal Hamiltonian $H=\sum_{x\in\bbZ}h_x$.
The local Hamiltonian $h_x$ acts nontrivially only on sites $y$ such that $|y-x|\le R$, and satisfies $\snorm{h_x}\le B$, where $R$ and $B$ are constants independent of $x$.
We assume that each $h_x$ is invariant under the $\ZZ$ transformation given by $(S^{(1)}_x,S^{(2)}_x,S^{(3)}_x)\to(S^{(1)}_x,-S^{(2)}_x,-S^{(3)}_x)$ and $(S^{(1)}_x,S^{(2)}_x,S^{(3)}_x)\to(-S^{(1)}_x,-S^{(2)}_x,S^{(3)}_x)$, or the time-reversal symmetry transformation given by $(S^{(1)}_x,S^{(2)}_x,S^{(3)}_x)\to(-S^{(1)}_x,-S^{(2)}_x,-S^{(3)}_x)$.
See section~\ref{s:SR} for details.
We also note that any system which is $SO(3)$ invariant is automatically invariant under this $\ZZ$ symmetry. This plays an important role in Appendix~\ref{app:Lie}, where we discuss the cases with compact Lie group symmetry.

Let us first assume that the model has translation invariance, i.e., $\bar{S}_x=\bar{S}$ and $h_{x+1}=\tau_1(h_{x})$  for all $x$, where $\tau_1$ is the translation operator.
Then the following was conjectured by  Chen, Gu, and Wen \cite{ChenGuWEn2011}, proved for MPS by Watanabe, Po, Vishwanath, and Zaletel \cite{WPVZ}, and was proved by Ogata and Tasaki \cite{OT1}:
\begin{C}
\label{C:trans}
If $\bar{S}$ is a half-odd integer, then it is never the case that the above translation-invariant model (with $\ZZ$ or time-reversal invariance) has a unique gapped ground state.
\end{C}
In \cite{OT1} the cases for $\ZZ$ symmetry and time-reversal symmetry were treated separately.
Here we prove a much more general and unified result, Theorem~\ref{T:trans} below, which corresponds to the original conjecture by Chen, Gu, and Wen \cite{ChenGuWEn2011}.

Let us next assume that the model is invariant under the reflection about the origin.
We assume the symmetry $\bar{S}_x=\bar{S}_{-x}$ and $h_{-x}=\calR(h_{x})$ for all $x\in\bbZ$, where $\calR$ denotes the reflection map.
Then our main result is the following Lieb-Schultz-Mattis type statement.
\begin{C}
\label{C:ref}
If $\bar{S}_0$ is a half-odd integer, then it is never the case that the above reflection-invariant model  (with $\ZZ$ or time-reversal invariance)  has a unique gapped ground state.
\end{C}
Note that, rather remarkably, the condition for the corollary contains only the spin quantum number $\bar{S}_0$ at the origin; $\bar{S}_x=\bar{S}_{-x}$ on other sites are arbitrary.
The above statement is a simple corollary of our general result, Theorem~\ref{T:ref}.
The corollary is reminiscent of the well-known fact, often called the Kramers degeneracy, that all the energy eigenvalues are inevitably even-fold degenerate in a system of a single half-odd-integral spin with a $\ZZ$ or time-reversal
invariant Hamiltonian.
See, e.g., \cite[Chapter 2]{TasakiBook}.

The statement of Corollary~\ref{C:ref} was discussed first by Fuji \cite[section~3.D]{Fuji}, and then in a more general context by Po, Watanabe, Jian, and Zaletel \cite{PWJZ}.
In fact the corollary shows that the model depicted in \cite[Figure~1(i)]{PWJZ} cannot have a unique gapped ground state, confirming their conjecture (restricted to quantum spin chains).
We also note that an earlier result in  \cite{HiranoKatsuraHatsugai} also suggests Corollary~\ref{C:ref}.

\section{Outline of the argument}
\label{s:outlie}

Before discussing the settings and results in detail, we recall the notions of projective representations of a symmetry group and the corresponding second group cohomology, and also give an informal account of the proofs of the main results.

\paragraph*{Projective representations and degree-2 cohomology classes}
Let $G$ be a finite group that describes on-site symmetry of the spin system.
We fix a homomorphism $\ua:G\to\{1,-1\}$,
which gives a decomposition\footnote{%
This decomposition  is known as a UA-decomposition.
See, e.g., \cite{Par,LS}. 
The same structure is also known as a Real structure on a group following Atiyah;
see, e.g., \cite[Chapter 2]{BG}.
} $G=G^+\sqcup G^-$ with $G^\pm\coloneqq\{g\in G\mid\ua(g)=\pm1\}$.
In the following we always consider the pair $(G,\ua)$ as the basic data
and denote it simply by $G$, leaving $\ua$ implicit.
In the main part of the paper, we assume $G$ is finite,
but we can generalize our theorems to compact Lie groups.
We will give a brief discussion on this generalized case in Appendix~\ref{app:Lie}.

Let $\calH$ be a Hilbert space.
The collection of operators $V(g)$ on $\calH$ with $g\in G$  is said to be a projective representation of $G$ if 
\begin{itemize}
\item $V(e)=\Id$,
\item $V(g)$ is unitary if $\ua(g)=1$ and antiunitary\footnote{%
A map $A:\calH\to\calH$ is said to be an antilinear operator if $A(\alpha u+\beta v)=\bar{\alpha}Au+\bar{\beta}Av$ for any $\alpha,\beta\in\bbC$ and  $u,v\in\calH$.
The adjoint $A^*$ of a bounded antilinear operator $A$ is the unique antilinear operator that satisfies $\sbkt{u,Av}=\overline{\sbkt{A^*u,v}}$ for any $u,v\in\calH$.
An antilinear operator $V$ that satisfies $VV^*=V^*V=\Id$ is said to be an antiunitary operator.
} if $\ua(g)=-1$,\footnote{%
Such an assignment of operators was called a co-representation by Wigner.
In our paper we call co-representations simply as representations, as this would not cause any confusions.
}
\item
 and
\eq
V(g)\,V(h)=\varphi(g,h)\,V(gh)
\lb{VVV}
\en
with $\varphi(g,h)\in\Uo=\{z\in\bbC\,\bigl|\,|z|=1\}$ for any $g,h\in G$.
\end{itemize}
From associativity $V(f)\{V(g)\,V(h)\}=\{V(f)\,V(g)\}V(h)$ and \rlb{VVV}, one finds that $\varphi$ must satisfy
\eq
\frac{{\overline{\varphi(g,h)}}^{\,\ua(f)}\varphi(f,gh)}{\varphi(f,g)\,\varphi(fg,h)}=1\ \text{for any $f,g,h\in G$,}
\lb{cocycle1}
\en
where we define $\bar{z}^{\,\ua}:=z$ if $\ua=1$ and $\bar{z}^{\,\ua}:=\bar{z}$ if $\ua=-1$.
We also see from \rlb{VVV} and $V(e)=\Id$ that
\eq
\varphi(g,e)=\varphi(e,g)=1\ \text{for any $g\in G$.}
\lb{cocycle2}
\en
In general a map $\varphi:G\times G\to\Uo$ that satisfies \rlb{cocycle1} and  \rlb{cocycle2} is called a 2-cocycle of $G$.
We define the product of two 2-cocycles as their point-wise product.
Then the set of all 2-cocycles of $G$ becomes an abelian group, which we denote as $Z^2(G,\Uu)$.

Suppose that there is another projective representation $\tilde{V}(\cdot)$ of $G$ with 2-cocycle $\tilde{\varphi}$, and it is related to $V(\cdot)$ by $V(g)=\psi(g)\,\tilde{V}(g)$ with $\psi(g)\in\Uo$ for any $g\in G$.
From \rlb{VVV} we see that the two 2-cocycles are related by
\eq
\varphi(g,h)=\frac{\psi(g)\,{\overline{\psi(h)}}^{\,\ua(g)}}{\psi(gh)}\,\tilde{\varphi}(g,h)\ \text{for any $g,h\in G$.}
\lb{coceq}
\en
This motivates us to define, in general, two 2-cocycles $\varphi$ and $\tilde{\varphi}$ related by \rlb{coceq} with some $\psi$ to be equivalent with each other.
We denote the set of corresponding equivalence classes of $Z^2(G,\Uu)$ as $\SC$.
The quotient set $\SC$ also becomes an abelian group, and is called the second group cohomology of $G$.\footnote{%
When  $\ua(g)=1$ for all $g\in G$, $\SC$ is written as $H^2(G,\Uo)$.
}
One can thus associate a unique element $\sigma$ of $\SC$ with any projective representation $V(\cdot)$ of $G$.
We say that $\sigma$ is the degree-2 cohomology class of the projective representation $V(\cdot)$.

Our theorems are meaningful when $\SC$ is nontrivial.
We have\footnote{%
We shall express $\SC$ as an additive group throughout the present paper.
} $\SC=\{0,1\}$ for the two important cases discussed in section~\ref{s:intro}, namely, $\ZZ$ or time-reversal symmetry.
See section~\ref{s:SR}.

\begin{figure}
\centerline{\includegraphics[width=9.5cm]{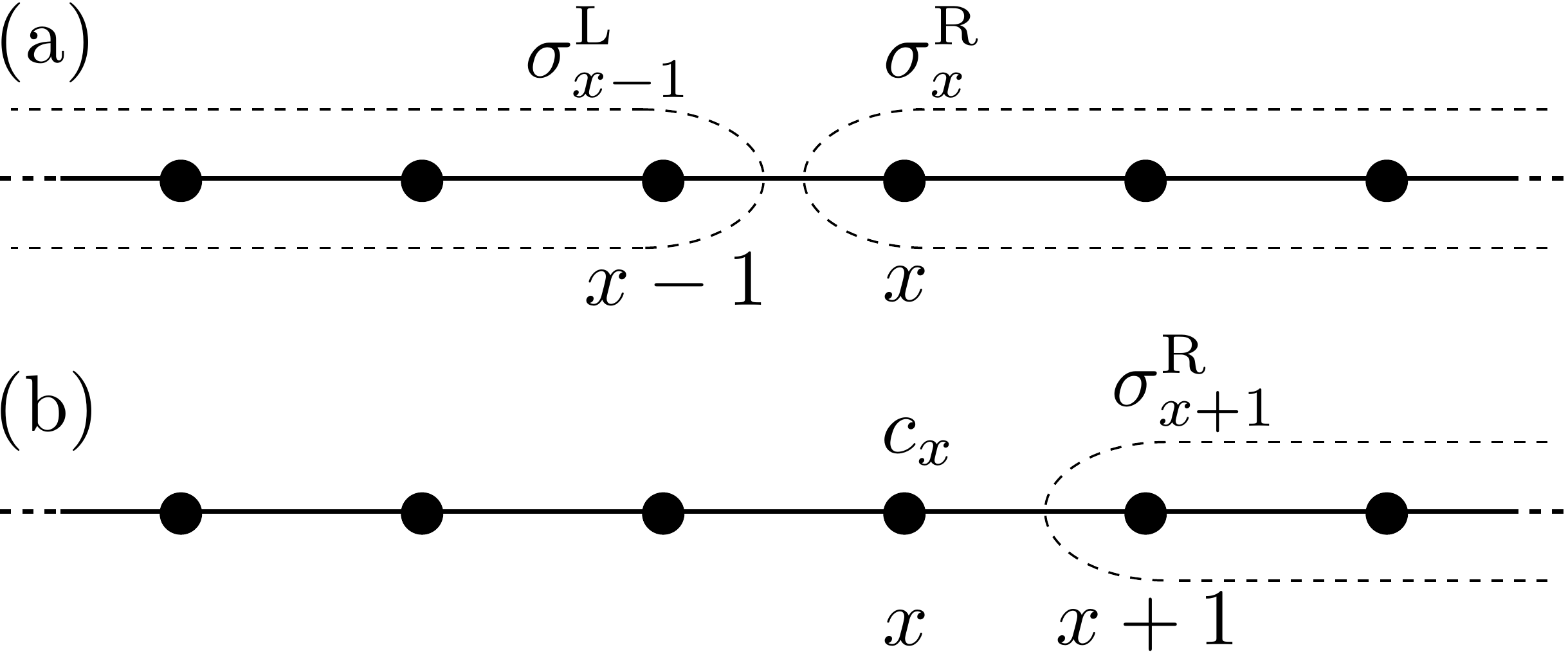}}
\caption[dummy]{
One can associate a unique index $\sigma\in\SC$ with the pure split state $\rho$ restricted onto a half-infinite chain.
(a)~The indices $\sL_{x-1}$ and $\sR_x$ describe the transformation properties of ``edges states'' that emerge when the infinite chain is decomposed into two half-infinite chains $\{\ldots,x-2,x-1\}$ and $\{x,x+1,\ldots\}$.
(b)~The half infinite chain $\{x,x+1,\ldots\}$ may be regarded as consisting of the site $x$ and $\{x+1,x+2,\ldots\}$.
We have the corresponding identity \rlb{2}, which is a key ingredient of the present work.
}
\label{f:chain}
\end{figure}

\paragraph*{``Edge states'' and the Lieb-Schultz-Mattis type theorems}
We consider a quantum spin system on the infinite chain $\bbZ$ with a certain symmetry group $G$, accompanied by a homomorphism $\ua$ giving the decomposition $G=G^+ \sqcup G^-$.
We assume that there is a projective representation of $G$ at each site $x$, and denote by $c_x\in\SC$ the corresponding degree-2 cohomology class.

We then take a pure state $\rho$ that is invariant under the global action of $G$ and also satisfies the property called the split property.
See Definition~\ref{D:split} below.
A unique ground state accompanied by a nonzero energy gap of the quantum spin chains described in section~\ref{s:intro} is an example.
See the end of section~\ref{s:proof1} for details.

Suppose that one decomposes the infinite chain $\bbZ$ into two half-infinite chains as $\bbZ=\{\ldots,x-2,x-1\}\cup\{x,x+1,\ldots\}$.
It was pointed out by Ogata \cite{Ogata2} that, by using notions from operator algebraic approaches to quantum spin systems, one can associate  a unique degree-2 cohomology class in $\SC$ with the state $\rho$ restricted onto each of the half-infinite chains.
We denote the degree-2 cohomology classes corresponding to the half-infinite chains $\{\ldots,x-2,x-1\}$ and $\{x,x+1,\ldots\}$ as $\sL_{x-1}$ and $\sR_x$, respectively.
See Figure~\ref{f:chain}~(a).
Physically speaking, $\sL_{x-1}$ and $\sR_x$ characterize the symmetry properties of ``edge states'' that emerge when the infinite chain is decomposed into two.
They correspond to the ``topological'' indices discussed intensively in the context of symmetry protected topological phases \cite{Perez-Garcia2008,PollmannTurnerBergOshikawa2010,ChenGuWEn2011,ZengChenZhouWenBOOK,TasakiBook}.

It was proved by Ogata \cite[Lemma 2.5]{Ogata2} that these indices satisfy
\eq
\sL_{x-1}+\sR_x=0.
\lb{1}
\en
The identity is natural if we recall that the two ``edge states'' emerge from a single pure state.
The main ingredient of the present work is the identity
\eq
\sR_x=c_x+\sR_{x+1},
\lb{2}
\en
which is proved in Lemma~\ref{L:main} below.
This relation is also natural since the half-infinite chain $\{x,x+1,\ldots\}$ may be regarded as consisting of a single site $x$ and the half-infinite chain $\{x+1,x+2,\ldots\}$.
Compare (a) and (b) of  Figure~\ref{f:chain}.

With the two identities \rlb{1} and \rlb{2}, we can easily prove Lieb-Schultz-Mattis type theorems that lead to Corollaries~\ref{C:trans} and \ref{C:ref}.
First assume that the state $\rho$ is translation  invariant.
Since we then have $\sR_x=\sR_{x+1}$, we find $c_x=0$ from \rlb{2}, i.e., the degree-2 cohomology class of the projective representation at each site must be trivial.
For $\ZZ$ or time-reversal symmetry, this means that the spin quantum number $\bar{S}$ is an integer (see section~\ref{s:SR}).
This implies the desired no-go statement, Corollary~\ref{C:trans}.
Next assume that $\rho$ is invariant under the reflection about the origin.
We then have $\sR_0=\sL_0$, which, with \rlb{1}, implies $\sR_0=-\sR_1$.
Substituting this into \rlb{2} with $x=0$, we find $c_0=2\sR_0$.
When  $\SC=\{0,1\}$ this is possible only when $c_0=0$.
We then get Corollary~\ref{C:ref}.

\section{Setting and main results}
\label{s:SR}
\paragraph*{C$^*$-algebras and split states}
We start by defining a general quantum spin system on the infinite chain $\bbZ$.
For each site\footnote{%
A site may be a collection of sites in the standard sense.
} $x\in\bbZ$ we associate a Hilbert space $\hilb_x\cong\bbC^{d_x}$ with dimension $d_x\in\bbN$. 
For a finite subset $\La\in\bbZ$, we define the algebra $\OA_\La$ of local observables as the set of all bounded operators on the Hilbert space $\bigotimes_{x\in\La}\hilb_x$.
For finite subsets $\La\subset\La'\subset\bbZ$, the algebra $\OA_\La$ is naturally embedded in $\OA_{\La'}$ by tensoring its elements by identity.
For any infinite subset $\Gamma\subset\bbZ$, we denote by $\OA_\Gamma$ the inductive limit of the collection of algebras $\OA_\La$ with $\La$ being an arbitrary finite subset of $\Gamma$.
The C$^*$-algebra of the whole chain is then denoted as $\OA=\OA_\bbZ$.
We also introduce the C$^*$-algebras for half-infinite chains by $\AL_x=\OA_{\{\ldots,x-1,x\}}$ and $\AR_x=\OA_{\{x,x+1,\ldots\}}$, where $x\in\bbZ$.
Note that $\AL_x$ and $\AR_x$ can naturally be regarded as subalgebras of $\OA$.

We define states on the C$^*$-algebras as usual.
The notion of split states is essential.
\begin{D}
\label{D:split}
Let $\rho$ be a pure state on $\OA$, and denote by $\rho^\rL_0$ and $\rho^\rR_1$ be the restrictions of $\rho$ onto the subalgebras $\AL_0$ and $\AR_1$, respectively.
We say that $\rho$ satisfies the split property if $\rho$ and  $\rho^\rL_0\otimes\rho^\rR_1$ are quasi-equivalent.
\end{D}
It is easily seen that one may replace  $\rho^\rL_0$ and $\rho^\rR_1$ in the above definition by  $\rho^\rL_x$ and $\rho^\rR_{x+1}$ with an arbitrary $x\in\bbZ$, which are the restrictions of $\rho$ onto $\AL_x$ and $\AR_{x+1}$, respectively.

In the main theorems, Theorems~\ref{T:trans} and \ref{T:ref}, we state necessary conditions for the existence of a pure split state.
They can be rephrased as no-go theorems for unique gapped ground states since a unique gapped ground state (of a model with short range interactions) is known to satisfy the split property\cite{Matsui2}.\footnote{%
It is also known that a state with area law entanglement satisfies the split property.
Thus our theorems may be interpreted as no-go theorems for area law states.
}
See the end of section~\ref{s:proof1}. 

\paragraph*{On-site symmetry}
We always consider a model with certain on-site symmetry.
Let $G$ be a finite group and fix a homomorphism $\ua:G\to\{1,-1\}$.
For each $x\in\bbZ$, we assume that there is an operator $v_x(g)$ on $\hilb_x$ which is unitary if $\ua(g)=1$ and antiunitary if $\ua(g)=-1$, and that $v_x(\cdot)$ gives a projective representation of the group $G$.
We denote by $c_x\in\SC$ the degree-2 cohomology class of the projective representation,
as explained in section \ref{s:outlie}.

We define the adjoint representation of $G$ by $\Xi(g)(A)=v_x(g)A(v_x(g))^*$ for $A\in\OA_{\{x\}}$.
One can uniquely extend $\Xi(g)$ to  \mbox{$*$-automorphisms} on $\OA$, $\AL_x$, and $\AR_x$.
The \mbox{$*$-automorphism} $\Xi(g)$ is linear if $\ua(g)=1$ and antilinear if $\ua(g)=-1$.
Note that $\Xi(\cdot)$ gives a genuine representation of $G$, i.e., $\Xi(g)\circ\Xi(h)=\Xi(gh)$ for any $g,h\in G$.
We say that a state $\rho$ on $\OA$ is $G$-invariant if $\rho(\Xi(g)(A))=\rho(A)$ when $\ua(g)=1$ and
$\rho(\Xi(g)(A^{*}))=\rho(A)$ when $\ua(g)=-1$ for any $A\in\OA$.

\paragraph{Examples in section~\ref{s:intro} in this language} 
To see two examples discussed in section~\ref{s:intro}, we consider standard quantum spin systems on $\bbZ$.
The dimension of the local Hilbert space $\hilb_x$ is given by $d_x=2\bar{S}_x+1$, where $\bar{S}_x\in\bbN/2$ is the spin quantum number at site $x\in\bbZ$.
For a finite subset $\La\subset\bbZ$, the algebra $\OA_\La$ consists of polynomials of spin operators $S^{(\nu)}_x$ with $\nu=1,2,3$ and $x\in\La$;
this is because  the operators $S^{(1,2,3)}_x$ generate the algebra $\OA_{\{x\}}$.

To formulate $\ZZ$ transformation, we set\footnote{%
The multiplication rule is $a_1a_2=a_3$, $a_2a_3=a_1$, $a_3a_1=a_2$, and $(a_\nu)^2=e$ for $\nu=1,2,3$.
} $G=\ZZ=\{e,a_1,a_2,a_3\}$, and $\ua(g)=1$ for all $g\in G$.
Then the second cohomology group is $H^2(\ZZ,\Uo)=\{0,1\}$.
We define the projective representation on the local Hilbert space $\hilb_x$ by $v_x(e)=\Id$ and $v_x(a_\nu)=\exp[-i\pi S^{(\nu)}_x]$ for $\nu=1,2,3$.
The degree-2 cohomology class of the projective representation is $c_x=0$ if $\bar{S}_x$ is an integer, and $c_x=1$ if $\bar{S}_x$ is a half-odd-integer.
It is found that the corresponding adjoint representation satisfies
\eq
\Xi(a_\nu)(S^{(\mu)}_x)=\begin{cases}
S^{(\mu)}_x&\nu=\mu,\\
-S^{(\mu)}_x&\nu\ne\mu,
\end{cases}
\lb{ZZ}
\en
for any $x\in\bbZ$ and $\nu,\mu=1,2,3$.

To formulate time-reversal transformation, we set $G=\bbZ_2=\{e,a\}$ and $\ua(e)=1$, $\ua(a)=-1$.
Then one has $H^2(\bbZ_2,\Uu)=\{0,1\}$.
We define the projective representation on $\hilb_x$ by $v_x(e)=\Id$ and $v_x(a)=K\,\exp[-i\pi S^{(2)}_x]$, where $K$ is the complex conjugation map.\footnote{%
We here use the standard matrix representation of spin operators in which all the matrix entries of $S^{(2)}_x$ are pure imaginary.
See, e.g., \cite[section~2.3]{TasakiBook} for details.
}
The degree-2 cohomology class of the projective representation is again $c_x=0$ if $\bar{S}_x$ is an integer, and $c_x=1$ if $\bar{S}_x$ is a half-odd-integer.
The corresponding adjoint representation gives $\Xi(a)(S^{(\nu)}_x)=-S^{(\nu)}_x$ for any $x\in\bbZ$ and $\nu=1,2,3$.

\paragraph*{Translation symmetry}
We shall describe a class of models with translation symmetry.
Take $d_x=d$ for all $x\in\bbZ$.
We can then regard each $\hilb_x$ as a copy of a single Hilbert space $\hilb_0\cong\bbC^d$.
For any $x,y\in\bbZ$ and $A\in\OA_{\{x\}}$, we denote by $\tau_y(A)\in\OA_{\{x+y\}}$ the identical copy of $A$ in $\OA_{x+y}$.
We also assume that the on-site symmetry transformation is chosen so that $v_{x+y}(g)$ is an identical copy of $v_x(g)$ on $\hilb_{x+y}$.
We then have $c_x=c\in\SC$ for all $x\in\bbZ$.

The translation $\tau_y$ automatically extends to a liner \mbox{$*$-automorphism} on $\OA$.
We say that a state $\rho$ on $\OA$ is translation invariant if $\rho(\tau_y(A))=\rho(A)$ for any $y\in\bbZ$ and $A\in\OA$.
Here we  prove the following theorem,
which contains two theorems proved in \cite{OT1} and summarized as Corollary~\ref{C:trans} as special cases.
\begin{T}\label{T:trans}
Consider a system with translation symmetry, and let $\rho$ be a pure split state that is $G$-invariant and translation invariant.
Then one inevitably has $c=0$.
\end{T}

\paragraph*{Reflection symmetry}
We consider another class of models that are invariant under reflection about the origin of the chain (but not necessarily invariant under translation).
Assume that the local dimensions satisfy $d_x=d_{-x}$ for all $x\in\bbZ$.
We can then take the local Hilbert spaces $\hilb_{x}$ and $\hilb_{-x}$ to be identical.
We assume that, for each $x\in\bbZ$, there is a linear \mbox{$*$-automorphism} $\calR_x:\OA_{\{x\}}\to\OA_{\{-x\}}$ such that $\calR_{-x}\circ\calR_{x}=\Id$.
We also assume that $v_{-x}(g)$ is an identical copy on $\hilb_{-x}$ of $v_x(g)$.

From $\calR_x$ with all $x\in\bbZ$, one can define a linear \mbox{$*$-automorphism} $\calR$ on $\OA$ such that $\calR(A)=\calR_x(A)$ for $A\in \OA_{\{x\}}$.
We say that a state $\rho$ on $\OA$ is reflection invariant if $\rho(\calR(A))=\rho(A)$ for any  $A\in\OA$.
Then the following is a general form of our new Lieb-Schultz-Mattis type theorem.
\begin{T}\label{T:ref}
Consider a system with reflection symmetry, and let $\rho$ be a pure split state that is $G$-invariant and reflection invariant.
Then one inevitably has $c_0=2c$ with some $c\in\SC$.
\end{T}
As we have already noted, the conclusion of the theorem implies $c_0=0$ when $\SC=\{0,1\}$ as in
the two models with $\ZZ$ or time-reversal symmetry discussed in section~\ref{s:intro}.
We also note that, by using the original idea of Lieb, Schultz, and Mattis, one can prove a similar (but different) theorem for a class of quantum spin chains with $\Uo\rtimes\bbZ_2$ symmetry.
See Appendix~\ref{app-U(1)}.

\begin{figure}
\centering
\includegraphics[width=.5\textwidth]{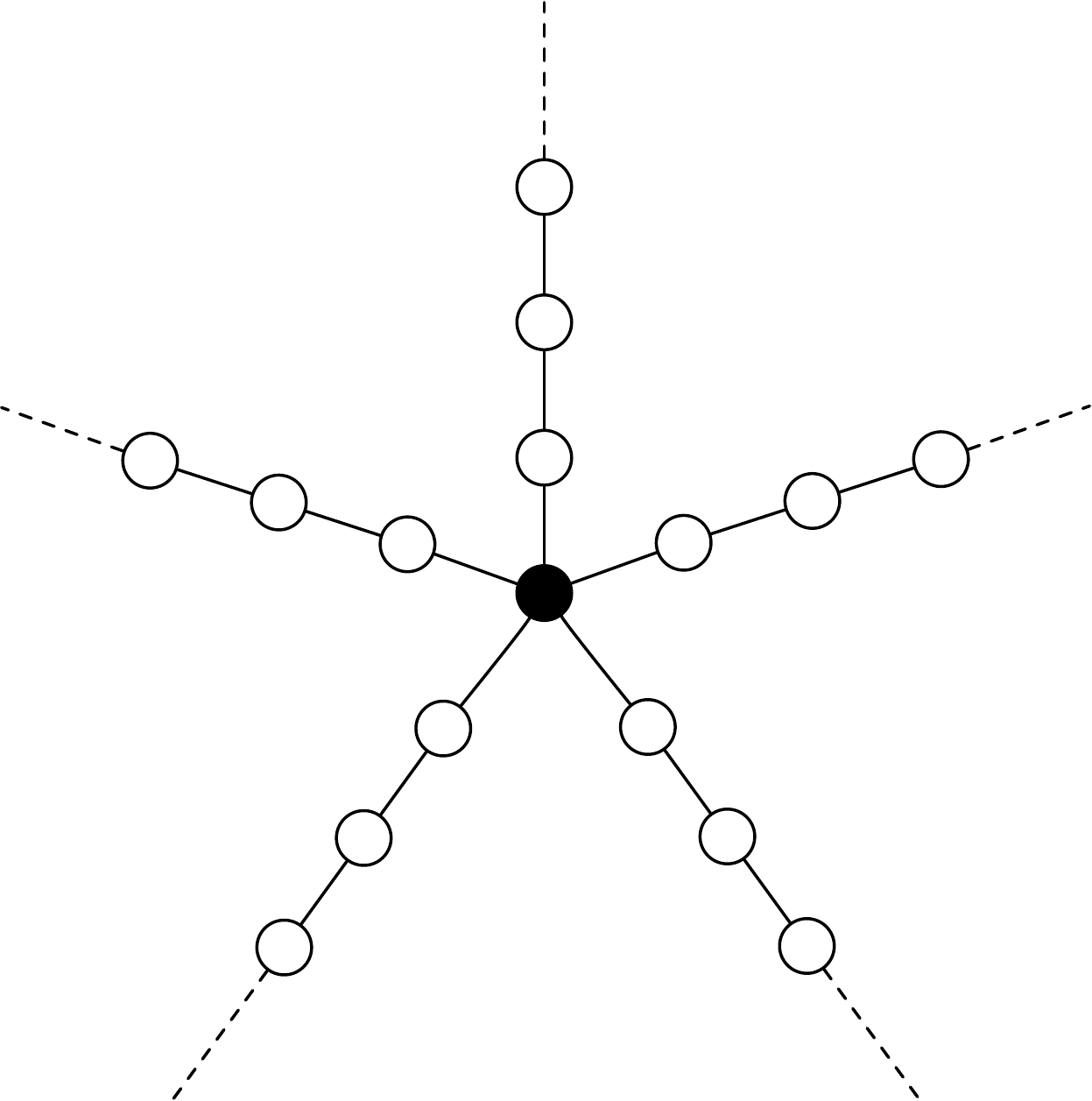}
\caption{%
The lattice $\La_m$ consists of the central site (black dot) and $m$ semi-infinite chains attached to it. The figure is for $m=5$.
\label{fig:foo}}
\end{figure}

There is a generalization of Theorem~\ref{T:ref} to $\bbZ_m$ invariant quantum spin system on the lattice $\Lambda_m\coloneqq\{o\}\cup\{(i,x)\mid i=1,\ldots,m, x=1,2,\ldots\}$, which consists of the central site $o$ and $m$ semi-infinite chains attached to it.
See Fig.~\ref{fig:foo} for the case $m=5$. 
We associate with the central site $o$ a Hilbert space $\hilb_o$ and a projective representation $v_o(\cdot)$ of $G$.
We impose $\mathbb{Z}_m$ symmetry by requiring that, for each $x=1,2,\ldots$, the Hilbert space $\hilb_{(i,x)}$ associated with site $(i,x)$ for $i=1,\ldots,m$ is an identical copy of a single Hilbert space $\hilb_x$, and also that the corresponding projective representation $v_{(i,x)}(\cdot)$ is identical to  $v_x(\cdot)$.
We consider the transformation that shifts the chain-index as $i\to i+1$, where we identify $m+1$ with 1.
This defines $\mathbb{Z}_m$ symmetry.

To define the split property for this system,
we note that the quantum spin system on $\Lambda_m$ can be regarded as a quantum spin chain by identifying the central site $o\in\Lambda_m$ with the origin $0\in\bbZ$, the site $(1,x)\in\Lambda_m$ with $x\in\bbZ$, and the collection of $m-1$ sites $(2,x),\ldots,(m,x)$ with $-x\in\bbZ$.
We then say that a $\bbZ_m$-invariant state $\rho$ on $\Lambda_m$ satisfies the split property if the  state on $\bbZ$ obtained by the above identification satisfies the split property.

\begin{T}\label{T:Zm}
Consider the quantum spin system on $\La_m$, and let $\rho$ be a $\bbZ_m$ and $G$-invariant pure state that satisfies the split property.
Then one inevitably has $c_o=mc$ with some $c\in\SC$.
\end{T}
Clearly Theorem \ref{T:ref} is a special case of Theorem \ref{T:Zm} for $m=2$.

\section{Indices for half-infinite chains and the proofs of theorems}
\label{s:proof1}
Let us discuss key ingredients of the present work, and prove the theorems and the corollaries.
Throughout the present section, we assume that $\rho$ is a $G$-invariant pure split state.

\paragraph*{Definition of indices}
We first follow Ogata \cite{OgataZ2}, and define indices $\sigma^\rL_x,\sigma^\rR_x\in\SC$ associated with the state $\rho$ restricted on the half-infinite chains.
See Figure~\ref{f:chain}~(a).

Let us fix $x\in\bbZ$ and $\rD\in\{\rL,\rR\}$.
Let $\rho^\rD_x$ be the restriction of $\rho$ onto the subalgebra $\AD_x$.
Let $(\calH^\rD_x,\uppi^\rD_x,\Omega^\rD_x)$ be the GNS triple corresponding to $\AD_x$ and $\rho^\rD_x$.
Since $\rho^\rD_x$ is $G$-invariant, one can use a standard argument (see, e.g.~\cite[Section~2.3.3]{BR1}) to define a \mbox{$*$-automorphism} $\hXi^\rD_x(g)$ on $\uppi^\rD_x(\AD_x)''$ that satisfies
\eq
\bigl(\hXi^\rD_x(g)\circ\uppi^\rD_x\bigr)(A)=\bigl(\uppi^\rD_x\circ\Xi(g)\bigr)(A),
\en
for any $A\in\AD_x$ and $g\in G$.
Again $\hXi^\rD_x(g)$ is linear if $\ua(g)=1$ and antilinear if $\ua(g)=-1$.

From the split property of $\rho$, it follows that $\uppi^\rD_x(\AD_x)''$ is a type-I factor, and hence is isomorphic to $B(\calK^\rD_x)$, the set of all bounded operators on a certain Hilbert space $\calK^\rD_x$.
Let us denote by $\iota^\rD_x:\uppi^\rD_x(\AD_x)''\to B(\calK^\rD_x)$ the corresponding \mbox{$*$-isomorphism}.
The space $\calK^\rD_x$ may be regarded as an effective Hilbert space that describes the states on the half-infinite chain that are close to $\rho^\rD_x$.

Combining the above, we get, for each $g\in G$, a \mbox{$*$-automorphism} $\iota^\rD_x\circ\hXi^\rD_x(g)\circ(\iota^\rD_x)^{-1}$ on $B(\calK^\rD_x)$.
It is linear if $\ua(g)=1$ and antilinear if $\ua(g)=-1$.
Then it follows from Wigner's theorem that there is an operator $V^\rD_x(g)$ on $\calK^\rD_x$ such that
\eq
\bigl(\iota^\rD_x\circ\hXi^\rD_x(g)\circ(\iota^\rD_x)^{-1}\bigr)(X)=V^\rD_x(g)\,X\,\bigl(V^\rD_x(g)\bigr)^*,
\lb{VXV}
\en
for any $X\in B(\calK^\rD_x)$.
The operator $V^\rD_x(g)$ is unitary if $\ua(g)=1$ and is antiunitary if $\ua(g)=-1$.
Clearly $V^\rD_x(\cdot)$ gives a projective representation of $G$.
Wigner's theorem also guarantees that the degree-2 cohomology class of the projective representation, which we denote as $\sD_x\in\SC$, is independent of the choice of $\calK^\rD_x$, $\iota^\rD_x$ or $V^\rD_x(\cdot)$.

\paragraph*{Properties of the indices}
The following basic property of the indices was proved by Ogata and plays an important role in the present work.
See Figure~\ref{f:chain}~(a).
\begin{Le}[{{=\cite[Lemma~2.5]{Ogata2} }}]
\label{L:LR0}
Let $\rho$ be a pure split state that is $G$-invariant.
Then the indices defined above satisfy $\sL_{x-1}+\sR_x=0$ for any $x\in\bbZ$.
\end{Le}

The most important ingredient of the present work is the following lemma, which relates the indices $\sD_x$ to the degree-2 cohomology class $c_x$ of the on-site projective representation of $G$ at site $x$.
See Figure~\ref{f:chain}~(b).
\begin{Le}
\label{L:main}
Let $\rho$ be a pure split state that is $G$-invariant.
Then the indices defined above satisfy
\eq
\sR_x=c_x+\sR_{x+1},\quad
\sL_x=c_x+\sL_{x-1}.
\lb{s=c+s}
\en
\end{Le}

The following two lemmas state invariance properties that follow from the assumed symmetry.
\begin{Le}
\label{L:trans}
Consider a system with translation symmetry, and let $\rho$ be a pure split state that is  $G$-invariant and translation invariant.
Then $\sL_x$ and $\sR_x$ are independent of $x$.
\end{Le}

\begin{Le}
\label{L:ref}
Consider a system with reflection symmetry, and let $\rho$ be a pure split state that is  $G$-invariant and reflection invariant.
Then one has $\sL_x=\sR_{-x}$ for any $x\in\bbZ$.
\end{Le}

We shall prove Lemmas~\ref{L:main}, \ref{L:trans}, and \ref{L:ref} in section~\ref{s:proofofLemma}.

\paragraph*{Proof of the theorems}
\label{s:proof2}
Let us prove the theorems, assuming Lemmas~\ref{L:LR0}, \ref{L:main}, \ref{L:trans}, and \ref{L:ref}.
Our strategy was already described in section~\ref{s:outlie}.

Assume that the state is translation invariant.
Then, since Lemma~\ref{L:trans} implies $\sR_x=\sR_{x+1}$, we readily find from \rlb{s=c+s} that $c_x=0$ for any $x$.
Theorem~\ref{T:trans} has been proved.

Assume that the state is reflection invariant.
We see from Lemma~\ref{L:ref} that $\sR_0=\sL_0$.
Noting that Lemma~\ref{L:LR0} implies $\sL_0+\sR_1=0$, we have $\sR_1=-\sR_0$.
We then see from \rlb{s=c+s}, in particular $\sR_0=c_0+\sR_1$, that $c_0=2\sR_0$.
Theorem~\ref{T:ref} has been proved.

Theorem~\ref{T:Zm} can be proved analogously by focusing on the corresponding state on the chain.
One only needs to note that Lemmas~\ref{L:main} and \ref{L:LR0} imply $\sR_0=c_0+\sR_1$ and $\sR_0=-\sL_{-1}$, respectively, and the $\bbZ_m$ symmetry implies $\sL_{-1}=(m-1)\sR_1$.
It then follows that $c_0=-m\sR_1$. \qedm

\paragraph*{Proof of the corollaries}
Consider a quantum spin system described in section~\ref{s:intro} and assume that the model has a unique gapped ground state.
(See, e.g., \cite{OT1} or \cite[Appendix~A.7]{TasakiBook} for a precise definition of unique gapped ground states.)
By using Hastings' result on the area law \cite{Hastings2007}, Matsui \cite{Matsui2} proved that such a ground state satisfies the split property.
Then, by noting that a unique ground state has the same symmetry as the Hamiltonian, we get Corollaries~\ref{C:trans} and \ref{C:ref} from Theorems~\ref{T:trans} and \ref{T:ref}, respectively.
It is also clear that a unique gapped ground state of the model on $\La_m$ (treated in Theorem~\ref{T:Zm}) satisfies the split property. \qedm

\section{Proof of Lemmas}
\label{s:proofofLemma}
\noindent\underline{Proof of Lemma~\ref{L:main}}:
Because of Lemma~\ref{L:LR0}, it suffices to prove one of the two relations in \rlb{s=c+s}.
We can also set $x=0$ without losing generality.
Our goal is thus to prove $\sR_0=c_0+\sR_1$.

We claim that there is a \mbox{$*$-isomorphism}
$\beta: \uppi^\rR_0(\AR_{1})''\to \uppi^\rR_1(\AR_{1})''$
such that 
\eq
\beta\circ\uppi^\rR_0(\Id_{0}\otimes B)
=\uppi^\rR_1(B),\quad B\in \AR_{1}.
\lb{bbc}
\en
To see this, let $\tilde H$ be the norm-closure of the subspace $\uppi^\rR_0(\AR_{1})\,\Omega^\rR_0$
of $\calH^\rR_0$,
and $p$ the orthogonal projection onto $\tilde H$.
Then $p\in \uppi^\rR_0(\AR_{1})'$ and $\tilde \uppi^\rR_0( B)\coloneqq\uppi^\rR_0( \Id_{0}\otimes B)\,p$, $B\in \AR_{1}$
defines a \mbox{$*$-representation} of $\AR_{1}$ on $\tilde H$.
(See the proof of \cite[Lemma 2.4.14]{BR1}.)
By definition of $\tilde H$, $\Omega^\rR_0$ is cyclic for  $\uppi^\rR_0(\AR_{1})$ in $\tilde H$, and
$(\tilde H, \tilde \uppi^\rR_0, \Omega^\rR_0)$ is a GNS triple of $\rho^\rR_1=\rho\vert_{\AR_{1}}$.
Namely we may regard $(\tilde H, \tilde \uppi^\rR_0, \Omega^\rR_0)=(\calH^\rR_1,\uppi^\rR_1,\Omega^\rR_1)$.
We define $\beta$ by
\eq
\beta(a)\coloneqq ap,\quad a\in \uppi^\rR_0(\AR_{1})''.
\en
Because of $p\in \uppi^\rR_0(\AR_{1})'$ and the definition of $\tilde \uppi^\rR_0=\uppi^\rR_1$,
this is a \mbox{$*$-homomorphism} satisfying \rlb{bbc}.

It is clear from \rlb{bbc} that the range of $\beta$ is in $ \uppi^\rR_1(\AR_{1})''$.
The homomorphism $\beta$ is actually onto $ \uppi^\rR_1(\AR_{1})''$.
Indeed, by the Kaplansky density Theorem, for any $b\in  \uppi^\rR_1(\AR_{1})''$, there is a bounded net $\{B_{\alpha}\}_{\alpha}$ in $\AR_{1}$
such that $\sw\lim_{\alpha}\uppi^\rR_1(B_{\alpha})=b$.
We then have a bounded net $\{\uppi^\rR_0( \Id\otimes B_{\alpha})\}_{\alpha}$ in $\uppi^\rR_0(\AR_{1})''$.
Because of the $\sw$ compactness of the unit ball of 
$\uppi^\rR_0(\AR_{1})''$, we may take a $\sw$ convergent subnet i.e., $\{\uppi^\rR_0( \Id\otimes B_{\alpha'})\}
_{\alpha'}$.
We use it to define $a$ by $\sw\lim_{\alpha'}\uppi^\rR_0(\Id\otimes B_{\alpha'})\eqqcolon a\in \uppi^\rR_0(\AR_{1})''$.
Then we obtain
\begin{align}
\beta(a)=ap=\sw\lim_{\alpha'}\uppi^\rR_0(\Id\otimes B_{\alpha'})\,p
=\sw\lim_{\alpha'}\uppi^\rR_1( B_{\alpha'})=b.
\end{align}
This proves the surjectivity.

To see that $\beta$ is injective, suppose $a\in \uppi^\rR_0(\AR_{1})''$ satisfies $\beta(a)=0$.
Then because $a\in  \uppi^\rR_0(\Az)'$, we have
\eqa
a\,\uppi^\rR_0(A\otimes B)\,\Omega^\rR_0
&=a\,\uppi^\rR_0(A\otimes \Id)\,\uppi^\rR_0(\Id\otimes B)\,\Omega^\rR_0
=\uppi^\rR_0(A\otimes \Id)\, a\,\uppi^\rR_0(\Id\otimes B)\,\Omega^\rR_0
\nl&=\uppi^\rR_0(A\otimes \Id) \,ap\,\uppi^\rR_0(\Id\otimes B)\,\Omega^\rR_0=0,
\ena
for all $A\in\Az$ and $B\in \AR_{1}$.
As vectors of the form $\uppi^\rR_0(A\otimes B)\,\Omega^\rR_0$ span
$\calH^\rR_0$, this means $a=0$. Hence $\beta$ is injective.

Take an arbitrary orthonormal basis $\{\ket{\psi_j}\}_{j=1,\ldots,d_0}$ of the local Hilbert space $\hilb_0$, and denote by $e_{j,k}=\ket{\psi_j}\bra{\psi_k}$ the corresponding matrix unit in $\Az$.

Recall that $\iota^\rR_0\circ\uppi^\rR_0$ is an irreducible representation of $\AR_0$ on $\calK^\rR_0$.
We define $\calN=(\iota^\rR_0\circ\uppi^\rR_0(e_{1,1}))\,\calK^\rR_0$, which is intuitively interpreted as the effective Hilbert space (for the half-infinite chain $\{0,1,\ldots\}$) with the spin at $x=0$ ``frozen'' into the state $\ket{\psi_1}$.
We then set $\tcK^\rR_0=\hilb_0\otimes\calN$, where we have ``supplied'' the missing spin.

We now construct \mbox{$*$-isomorphisms} $\Theta_0:\uppi^\rR_0(\AR_0)''\to B(\tcK^\rR_0)$ and $\Theta_{1}: \uppi^\rR_1(\AR_{1})''\to \calB(\calN)$. 
We start with the construction of $\Theta_0$.
Let us define an operator $U:\calK^\rR_0\to\tcK^\rR_0$ by
\eq
U\xi=\sum_{j=1}^{d_0}\ket{\psi_j}\otimes\bigl\{
(\iota^\rR_0\circ\uppi^\rR_0(e_{1,j}))\,\xi
\bigr\}.
\lb{Udef}
\en
One finds by inspection that the action of $U^*$ is given by
\eq
U^*\sum_{j=1}^{d_0}\ket{\psi_j}\otimes\eta_j=\sum_{j=1}^{d_0}(\iota^\rR_0\circ\uppi^\rR_0(e_{j,1}))\,\eta_j,
\lb{Ustar}
\en
where we wrote an arbitrary element of $\tcK^\rR_0$ in the form $\sum_{j=1}^{d_0}\ket{\psi_j}\otimes\eta_j$.
It can be easily checked that $U$ is unitary.
By using $U$, we define a \mbox{$*$-isomorphism} $\Theta_0:\uppi^\rR_0(\AR_0)''\to B(\tcK^\rR_0)$ by
\eq
\Theta_0(Y)=U\,\iota_0^\rR(Y)\,U^*,
\lb{Theta0}
\en
for $Y\in\uppi^\rR_0(\AR_0)''$.

We next construct  $\Theta_1$.
Let $A\in\Az$ and $B\in\AR_1$.
For any $\sum_{j=1}^{d_0}\ket{\psi_j}\otimes\eta_j\in\tcK^\rR_0$ we observe that
\begingroup\allowdisplaybreaks
\eqa
\bigl(\Theta_0\circ\uppi^\rR_0(A\otimes B)\bigr)\sum_{j=1}^{d_0}\ket{\psi_j}\otimes\eta_j&=
U\bigl(\iota_0^\rR\circ\uppi^\rR_0(A\otimes B)\bigr)U^*\sum_{j=1}^{d_0}\ket{\psi_j}\otimes\eta_j
\nl&=\sum_{j=1}^{d_0}U\,\bigl(\iota_0^\rR\circ\uppi^\rR_0(A\,e_{j,1}\otimes B)\bigr)\,\eta_j
\nl&=\sum_{i,j=1}^{d_0}\ket{\psi_i}\otimes\bigl\{\bigl(\iota_0^\rR\circ\uppi^\rR_0(e_{1,i}\,A\,e_{j,1}\otimes B)\bigr)\,\eta_j\bigr\}
\nl&=\sum_{i,j=1}^{d_0}\ket{\psi_i}\bra{\psi_i}A\ket{\psi_j}\otimes\bigl\{\bigl(\iota_0^\rR\circ\uppi^\rR_0(e_{1,1}\otimes B)\bigr)\,\eta_j\bigr\}
\nl&=\sum_{j=1}^{d_0}A\ket{\psi_j}\otimes\bigl\{\bigl(\iota_0^\rR\circ\uppi^\rR_0(e_{1,1}\otimes B)\bigr)\,\eta_j\bigr\},
\ena
\endgroup
where $\Id_0$ is the identity in $\Az$.
Let $\gamma$ be a \mbox{$*$-representation} of $\AR_{1}$ on $\calN$ defined by
\begin{align}
\gamma(B)\coloneqq
\iota_0^\rR\circ\uppi^\rR_0(e_{1,1}\otimes B),\quad
B\in \AR_{1}.
\end{align}
From above, we have
\begin{align}
\Theta_0\circ\uppi^\rR_0(A\otimes B)
=A\otimes\gamma(B),\quad A\in \Az,\quad B\in \AR_{1}.
\end{align}
Because $\Theta_{0}$ is a \mbox{$*$-isomorphism}, elements with these form in $\calB(\tcK^\rR_0)$ span a $\sw$ dense subspace of
$\Theta_{0}(\uppi^\rR_0(\AR_{0})'')=\calB(\tcK^\rR_0)$.
From this, we conclude that the commutant $\gamma(\AR_{1})'$ of $\gamma(\AR_{1})$
is trivial, because otherwise $\Theta_{0}(\uppi^\rR_0(\AR_{0})'')=\calB(\tcK^\rR_0)$
would have a non-trivial commutant.
Therefore, $\gamma$ is irreducible.
(See \cite[Section 2.3]{BR1}.)

Let $j: \calB(\calN)\to \bbC\Id_0\otimes \calB(\calN)$ be the \mbox{$*$-isomorphism}
given by $j(x)\coloneqq\Id_0\otimes x$, for $x\in \calB(\calN)$.
From above, $j^{-1}\circ\Theta_{0}\circ \beta^{-1}$ is well-defined on $(\uppi^\rR_1(\AR_{1}))''$ and we get
\begin{align}
j^{-1}\circ\Theta_{0}\circ \beta^{-1}\circ\uppi^\rR_1(B)=\gamma(B),\quad B\in \AR_{1}.
\end{align}
Set $\Theta_{1}\coloneqq j^{-1}\circ\Theta_{0}\circ \beta^{-1}: \uppi^\rR_1(\AR_{1})''\to \calB(\calN)$.
Again by the Kaplansky density theorem and the irreducibility of $\gamma$, 
$\Theta_{1}$ is a \mbox{$*$-isomorphism}.

Therefore, we may regard $\calN=\calK^\rR_1$.
By the definition, we have
\begin{align}
\gamma(B)=\Theta_{1}\circ \uppi^\rR_1(B),\quad B\in \AR_{1}.
\end{align}
We can then repeat the construction in \rlb{VXV} by using $\calN$ and $\Theta_1$ to have
\eq
\bigl(\Theta_1\circ\hXi^\rR_1(g)\circ\Theta_1^{-1}\bigr)(X)=\tV_1(g)\,X\,\bigl(\tV_1(g)\bigr)^*,
\lb{tVXV}
\en
for any $g\in G$ and $X\in B(\calN)$.
$\tV_1(g)$ is unitary if $\ua(g)=1$ and antiunitary if $\ua(g)=-1$.
We see that $\tV_1(\cdot)$ gives a projective representation of $G$ on $\calN$.
The degree-2 cohomology class of $\tV_1(\cdot)$ is $\sR_1$ by uniqueness that follows from Wigner's theorem.
Note that
\begin{align}
\tV_1(g)\,\gamma(B)\,\bigl(\tV_1(g)\bigr)^*
=\gamma(\Xi^\rR_1(g)(B)),\quad B\in\AR_{1}.
\end{align}

Let us finally define unitary or antiunitary operators on $\tcK^\rR_0=\hilb_0\otimes\calN$ by\footnote{%
Let $A_1$ and $A_2$ be antilinear operators on Hilbert spaces $\calH_1$ and $\calH_2$, respectively.
We denote by $A_1\otimes A_2$ the unique antilinear operator on $\calH_1\otimes\calH_2$ that satisfy $(A_1\otimes A_2)(v_1\otimes v_2)=(A_1v_1)\otimes(A_2v_2)$ for any $v_1\in\calH_1$ and $v_2\in\calH_2$.
}
\eq
\tV_0(g)=v_o(g)\otimes\tV_1(g).
\en
It is clear that $\tV_0(\cdot)$ forms a projective representation of $G$ with degree-2 cohomology class $c_0+\sR_1\in\SC$.
We shall show that the 2-cocycle associated to the projective representation  $\tV_0(\cdot)$ is equivalent to that given by $V^\rR_0(\cdot)$, and has the associated degree-2 cohomology class $\sR_0$.
This implies the desired identity $c_0+\sR_1=\sR_0$.

To confirm the claim, note for any $A\in\Az$ and $B\in\AR_1$ that
\begingroup\allowdisplaybreaks
\eqa
\tV_0(g)\bigl\{\Theta_0\circ\uppi^\rR_0(A\otimes B)\bigl\}(\tV_0(g))^*&=
\tV_0(g)\bigl(A\otimes\gamma(B))\bigr)(\tV_0(g))^*
\nl&=\bigl\{v_0(g)\,A\,(v_0(g))^*\bigr\}\otimes 
\bigl\{\tV_1(g)\bigl(\gamma(B)\bigr)(\tV_1(g))^*\bigr\}
\nl&=\Xi(g)(A)\otimes\bigl\{\bigl(\gamma\circ\Xi(g)\bigr)(B)\bigr\}
\nl&=\bigl(\Theta_0\circ\uppi^\rR_0\circ\Xi(g)\bigr)(A\otimes B)
\nl&=(\Theta_0\circ\hXi^\rR_0(g)\circ\uppi^\rR_0)(A\otimes B).
\ena
\endgroup
This implies
\eq
\bigl(\Theta_0\circ\hXi^\rR_0\circ(\Theta_0)^{-1}\bigr)(X)=\tV_0(g)\,X\,(\tV_0(g))^*,
\en
for any $X\in B(\tcK^\rR_0)$.
This should be compared with \rlb{VXV} with $x=0$ and $\rm D=R$.
Again from the uniqueness of the cohomology class, we see that the projective representation $\tV_0(\cdot)$ is characterized by $\sR_0$.~\qedm

\medskip
\noindent\underline{Proof of Lemmas~\ref{L:trans} and \ref{L:ref}}:
To prove the two lemmas in a unified manner,  consider two sub C$^*$-algebras $\OA_1$ and $\OA_2$
of $\OA$ related by a linear $*$-automorphism $\gamma$ on  $\OA$ such that $\gamma(\OA_2)=\OA_1$.
We assume that $\rho(\gamma(A))=\rho(A)$ for any $A\in\OA$, and that $\gamma\circ\Xi(g)=\Xi(g)\circ\gamma$.
In the context of Lemma~\ref{L:trans}, we set $\OA_1=\AR_x$, $\OA_2=\AR_y$ or $\OA_1=\AL_x$, $\OA_2=\AL_y$, and let $\gamma$ be the corresponding translation.
In the context of Lemma~\ref{L:ref}, we set $\OA_1=\AL_x$, $\OA_2=\AR_{-x}$, and let $\gamma$ be the reflection $\calR$.

Let $(\mathcal{H}, \pi, \Omega)$ be a GNS triple of $\rho\vert_{\OA_1}$, and 
$\iota$ be a $*$-isomorphism from $\pi(\OA_1)''$ onto $B(\mathcal{K})$, for some
Hilbert space $\mathcal{K}$.
Assume that there is a projective representation $V(g)$ of $G$ on $\mathcal{K}$ such that 
$Ad V(g)\circ\iota\circ\pi=
\iota\circ\pi\circ\Xi(g)$.
By the $\gamma$-invariance of $\rho$,  $(\mathcal{H}, \pi\circ\gamma , \Omega)$ 
is a GNS triple of 
$\rho\circ\gamma\vert_{\OA_2}=\rho\vert_{\OA_2}$.
Then, $\iota$ is a $*$-isomorphism from 
$\pi\circ\gamma(\OA_2)''=\pi(\OA_1)''$ onto $B(\mathcal{K})$,
and we have
\begin{align}
Ad V(g)\circ\iota\circ\pi\circ \gamma(A)
=\iota\circ\pi\circ\gamma\circ\Xi(g)(A), A\in \OA_2.
\end{align}
This means that $V(g)$ plays a role of $V^\rD_x(g)$
in \rlb{VXV} and completes the proof. \qedm



\section{Discussion}
\label{s:discussin}
We have developed a general method for proving Lieb-Schultz-Mattis type no-go theorems for quantum spin chains with on-site symmetry.
Our method makes use of the \emph{topological} indices that characterize projective representations of the symmetry that emerge at the edges of half-infinite chains.
In order to define meaningful indices in a mathematically rigorous manner, it was essential to follow \cite{OgataZ2} and  introduce the effective Hilbert space $\calK^\rD_x$ for the half-infinite chain through the von Neumann algebra $\uppi^\rD_x(\AD_x)''$.

By using this method we proved a general theorem, Theorem~\ref{T:trans}, for translation invariant models with on-site symmetry, which is a fully general and rigorous version of the conjecture stated by Chen, Gu, and Wen \cite{ChenGuWEn2011}.
We also  proved another general theorem, Theorem~\ref{T:ref}, which applies to models with on-site symmetry and the reflection invariance about the origin.
This statement previously appeared as a conjecture in a paper by Po, Watanabe, Jian, and Zaletel \cite{PWJZ}.

The reader might notice that, under the assumption that there is a pure split state, Theorem~\ref{T:ref} only poses a constraint $c_0=2c$ on the degree-2 cohomology class $c_0$ of the site at the origin, while Theorem~\ref{T:trans} completely determines the degree-2 cohomology class of all the sites as $c_x=0$.
This does not mean that Theorem~\ref{T:ref} is incomplete.
One can explicitly construct a reflection invariant pure split state with $c_0\ne0$, for example, in a model with on-site $\bbZ_3\times\bbZ_3$ symmetry, where the second group cohomology is given by $H^2(\bbZ_3\times\bbZ_3,\Uo)=\{0,1,2\}$.
See \cite{OT2}.

It is clear that Theorem~\ref{T:trans} can be readily extended to an on-site symmetric model that is invariant under translation followed by a global transformation (such as a spin rotation) which preserves the symmetry.
Likewise Theorem~\ref{T:ref} can be extended to an on-site symmetric model that is invariant under the reflection followed by a similar global transformation.
We do not, however, regard these statements as genuine extensions since these models may be transformed into translation invariant ones or reflection invariant ones.

Our 
Lieb-Schultz-Mattis type theorems for on-site symmetric quantum spin chains are obtained by detecting a nontrivial necessary condition for the existence of a pure split state that follows from basic properties of the degree-2 cohomology classes and the assumed geometric invariance of the model.
Recalling that translation and reflection are essentially the only nontrivial invariance of the infinite chain $\bbZ$, it is likely that 
Lieb-Schultz-Mattis type theorems for quantum bosonic spin chains with discrete on-site symmetry are essentially exhausted by our two theorems.

Similar 
Lieb-Schultz-Mattis type statements in two or higher dimensions have been discussed in the literature \cite{PTAV,WPVZ,PWJZ,Watanabe2018}.
It is quite challenging to see if these statements can be made into theorems by using similar operator algebraic techniques.

\medskip
{\small
It is a pleasure to thank Haruki Watanabe, Wojciech De Roeck, Chang-Tse Hsieh, and Hosho Katsura for useful discussions.
The present work was supported by JSPS Grants-in-Aid for Scientific Research nos.~16K05171 and 19K03534 (Y.O.),  16H06335 and 17H04837 (Y.T.),  and 16H02211 (H.T.). 
It was also supported by  JST CREST Grant Number JPMJCR19T2 (Y.O.).
}

\appendix
\section{Generalization to compact Lie groups}
\label{app:Lie}

In the main part of the paper we assumed that the on-site symmetry group $G$ is finite.
Here we outline how to generalize our setup to cover the case where the on-site symmetry group $G$ is a compact Lie group.
This generalization was first considered in \cite{DuivenvoordenQuella1}.

We note that the original theorem of Lieb, Schultz, Mattis given in \cite[Appendix B]{LSM} was for the $SU(2)$-symmetric Heisenberg chain, which can be thought of as an example of our general theorem when $G=SO(3)$, or $G=\ZZ\subset SO(3)$, as we used in section~\ref{s:intro}.
Lieb-Schultz-Mattis type theorems for compact Lie groups for more general compact groups have also been discussed in the literature, see \cite{YHO} and the references therein.

The main technical issue is that the definition of the group cohomology associated to a projective representation, given in section~\ref{s:outlie}, requires modifications when $G$ is a compact Lie group.
Just as we want a representation $V(g)$ of a Lie group $G$ to be a continuous map,
we need to impose some appropriate conditions on the cocycle $\varphi(g,h)$ as a function on $G\times G$.

We split the discussions in two cases, namely i) when the compact group $G$ is connected and the corresponding Lie algebra $\mathfrak{g}$ is semisimple, and then ii) when $G$ is a more general compact Lie group.
We give a proof in the case i); we only give an indication of a proof in the case ii).
The essential idea in the case i) is to find a suitable choice of finite subgroups $G_i$ of $G$  so that the projective representations of $G$ can be captured by those of $G_i$. 
This approach was already studied e.g.~in \cite{ElseBartlettDoherty,DuivenvoordenQuella2} when $G$ is a classical simple group except $Spin(4n)$.
Here we give a general construction applicable for arbitrary connected semisimple groups, based on a mathematical result \cite{Borel:1999bx}.
The main point in the case ii) is that a cohomology theory suitable for characterizing projective representations of continuous groups was already given by Mackey and Moore \cite{Mackey,Moore1,Moore2}.

\paragraph{When $G$ is a compact connected semisimple group}
In this case, let $\tilde G$ be the universal covering group of $G$ such that $G=\tilde G/H$ where $H$ is an abelian normal subgroup of $\tilde G$ so that we have the extension \begin{equation}
0\longrightarrow H\longrightarrow \tilde G \stackrel{p}{\longrightarrow} G \longrightarrow 0.
\end{equation}
We note that $H=\pi_1(G)$.

Let us consider a translation-invariant quantum spin chain where the Hilbert space $\hilb_x$ at each site is in a representation of $\tilde G$.
$\OA_x$ carries an adjoint action of $\tilde G$. 
Suppose that $\hilb_x$ as a representation of $H$ is a direct sum of a single irreducible representation $c$ of $H$.
Then $H$ acts trivially on $\OA_x$. 
Therefore $\OA_x$  is a representation of $G$, and therefore $\OA$ is a representation of $G$.
The version of theorem \ref{T:trans} in this setting is the following:
\begingroup
\renewcommand\thetheorem{\ref{T:trans}'}
\begin{T}
Consider a system with translation symmetry.
Suppose there is a pure split state $\rho$ that is $G$-invariant and translation invariant.
Then $c$ is the trivial representation of $H=\pi_1(G)$.
\end{T}
\endgroup
The proof of the theorem \ref{T:ref} for compact connected semisimple $G$ is entirely similar, so we omit it.
Our proof relies on the following lemma \ref{Le:trick}.

To motivate the context of the lemma, consider the case $G=SO(3)$, for which we have $\tilde G=SU(2)$ and $H=\mathbb{Z}_2$.
Recall that we used $G'=\ZZ\subset SO(3)$ as one of the main examples in the main part of the paper.
Let $\tilde G':=p^{-1}(G')$ so that $G'=\tilde G'/H$.
We then have the commutative diagram of extensions \begin{equation}
\begin{array}{ccccccc}
0 \longrightarrow& H&\longrightarrow& \tilde G& \longrightarrow&  G & \longrightarrow 0  \\
& \rotatebox{90}{$=$} & & \cup &&\cup \\
0 \longrightarrow& H&\longrightarrow& \tilde G'& \longrightarrow&  G' & \longrightarrow 0  
\end{array}.
\label{eq:com}
\end{equation}
An element $c\in \hat H$ then gives an extension of $G'$ by $\Uo$, therefore we have a homomorphism \begin{equation}
\iota: \hat H \to  H^2(G',\Uo).\label{eq:iota}
\end{equation}
For $G'=\ZZ$ this homomorphism $\iota$ is an isomorphism;
equivalently, $\tilde G'$ is the representation group of $G'=\ZZ$ in the sense of Schur,
i.e.~$\tilde G'$ is an extension of $G'$ by $H$ such that 
any projective representation of $G'$ is a genuine representation of $\tilde G'$ and $\hat H=H^2(G',U(1))$.
This allows us to reduce the LSM theorem for $G=SO(3)$ to the LSM theorem for $G'=\ZZ$.

This construction directly generalize when $\tilde G=SU(n)$, $H=\mathbb{Z}_n$ and $G=PSU(n)$.
In this case we consider two elements $a,b\in \tilde G$ given by \begin{align}
\label{standardab}
a&=\gamma\begin{pmatrix}
1 \\
& c  \\
&&c ^2 \\
&&&\ddots \\
&&&&c ^{n-2}\\
&&&&&c ^{n-1}\\
\end{pmatrix},&
b&=\gamma\begin{pmatrix}
0 & 1\\
&0&1\\
&&0&1\\
&&&\ddots &\ddots \\
&&&&0&1\\
1&&&&&0
\end{pmatrix},
\end{align}
where $c $ is a primitive $n$-th roots of unity and $\gamma^n=c ^{n(n-1)/2}$.
They satisfy $ab=c  ba$. 
Furthermore,  we have \begin{equation}
\label{abo}
a^n=b^n=\begin{cases}
1 & (n: \text{odd}),\\
c ^{n/2} & (n:\text{even}).
\end{cases}
\end{equation}
Then $\tilde G'=\langle a,b\rangle$ and $p(\tilde G')=G'=\mathbb{Z}_n\times\mathbb{Z}_n$  fit in the commutative diagram \eqref{eq:com} above.
We can now reduce the LSM theorem for $G=PSU(n)$ to that for $G'=\mathbb{Z}_n\times \mathbb{Z}_n$.

In general, two elements $a,b\in \tilde G$ in a  connected simply-connected Lie group such that $ab=c  ba$ where $c \in H$ are said to form an almost commuting pair, and all such pairs for an arbitrary $c$ were described in \cite{Schweigert:1996tg,Borel:1999bx,Kac:1999gw} for arbitrary Dynkin type. 
We will make use of the following lemma concerning almost commuting pairs:
\begin{Le}[\protect{=\cite[Corollary 4.2.1]{Borel:1999bx}}\ ]
\label{Le:abo}
Let $\tilde G$ be simply-connected semisimple compact group and $H$ be its center.
For any $c\in H$ of order $n$, there is an almost commuting pair $a,b\in G$ such that $ab=c ba$ satisfying \eqref{abo}. 
\end{Le}

From this lemma we derive another lemma given below:

\begin{Le}\label{Le:trick}
For any compact connected semisimple group $G$, there is a finite collection of finite subgroups $G'_i$ which fits in the commutative diagram
\begin{equation}
\begin{array}{ccccccc}
0 \longrightarrow& H&\longrightarrow& \tilde G& \longrightarrow&  G & \longrightarrow 0  \\
& \cup & & \cup &&\cup \\
0 \longrightarrow& H_i&\longrightarrow& \tilde G'_i& \longrightarrow&  G'_i & \longrightarrow 0  
\end{array}
\end{equation}
such that we have $\hat H_i \simeq H^2(G'_i,\Uo)$,
 $\prod H_i \simeq H$,
and therefore
$\hat H\simeq \prod \hat H_i \simeq \bigoplus  H^2(G'_i,\Uo)$.
\end{Le}

We provide the proof of the theorem first, and then that of the lemma.

\paragraph{Proof of the theorem \ref{T:trans}'}
A $G$-symmetric system is also $G'_i$-symmetric for each $i$.
We assumed that $\hilb_x$ is a direct sum of a single irreducible representation $c\in \hat H$.
This means that as a $G'_i$-symmetric system, the degree-2 cohomology class associated to $\hilb_x$ is given by $c_i\in H^2(G'_i,\Uo)$, where $c_i$ is the $i$-th direct sum component of $c$ in the decomposition $\hat H\simeq \bigoplus H^2(G'_i,\Uo)$.
We use the original Theorem~\ref{T:trans} and conclude that $c_i=0$ for each $i$.
We therefore conclude $c=0$.
\qedm

\paragraph{Proof of the lemma \ref{Le:trick}}
$H$ is a finite Abelian group, and therefore a product of cyclic groups, $H=\prod_i \mathbb{Z}_{k_i}$.
Let $\omega_i$ be the generator  of $\mathbb{Z}_{k_i}$.
From Lemma~\ref{Le:abo}, there is an almost commuting pair $a_i,b_i\in \tilde G$ for $\omega_i$.
We take $\tilde G'_i=\langle a,b\rangle$ and $G'_i=p(\tilde G'_i)$.
We have $G'_i=\mathbb{Z}_{k_i}\times \mathbb{Z}_{k_i}$ and $\tilde G'_i$ is its extension by $\mathbb{Z}_{k_i}$.
The resulting group is well-known to be  the representation group of $G'_i$.
\qedm

\paragraph{When $G$ is a general compact Lie group}
The discussion above is not  satisfactory, if one wants to consider the cases when $G$ is not necessarily connected, e.g.~$G=O(n)$.
In such cases it is not clear to the authors whether we can always choose finite subgroups $G'_i$ of $G$ so that the LSM theorem for $G$ can be deduced from the LSM theorems for $G'_i$ as we did above.

Instead, we can use a more general theory of projective representations of locally compact Lie groups $G$ developed in \cite{Mackey,Moore1,Moore2}.
There, the group cohomology $H^2(G,\Uu)$ is defined by placing the condition that the cochains $\varphi(g,h)$ on $G\times G$ is a Borel function.
When $G$ is connected and semisimple, the group cohomology defined in this manner is known to agree with $\widehat{\pi_1(G)}$ as used above \cite[Proposition 2.1]{Moore1}.

Another important theorem \cite[Theorem 2.2]{Mackey} for our purpose states that any continuous homomorphism $G\to PU(\mathcal{H})$
determines a unique class in $H^2(G,U(1))$, 
where $PU(\mathcal{H})=U(\mathcal{H})/U(1)$ is the projective unitary group for a Hilbert space $\mathcal{H}$.
Using this theorem instead of Wigner's theorem, 
we should be able to  attach the degree-2 classes $\sigma^{\rL}_x,\sigma^{\rR}_x \in H^2(G,\Uu)$ as in the finite group case, see the discussions around \eqref{e:VXV}.
The proof of the crucial lemma \ref{L:main} then goes mostly unchanged.

\paragraph{Comment on the case $G=U(1)$}
Our approach does not say anything nontrivial when $G=U(1)$, since $H^2(U(1),\Uo)=\{0\}$ when defined using the Borel functions.
This is unfortunate, since early extensions of the original theorem of LSM were mainly about relaxing the $SO(3)$ symmetry to its $U(1)$ subgroup. 
One important result in this direction was obtained in \cite{OYA}, where it was shown that there is no unique gapped ground state if the filling factor $\nu$ of a $U(1)$-symmetric system is not an integer. 

More recently, in \cite{CHR}, the various LSM theorems were given an interpretation in terms of the mixed anomaly between the on-site symmetry $G$ and the lattice translation symmetry $\mathbb{Z}$.
This is captured by an element in \begin{equation}
H^3(B(G\times \mathbb{Z}),\Uo)  / H^3(BG,\Uo) =H^2(BG,\Uo)
\end{equation} where $BG$ is the classifying space of $G$.
In many cases the group cohomology $H^2(BG,\Uo)$ defined in terms of the classifying space 
reduces to $H^2(G,\Uo)$ defined using cocycles on the group manifold. 
When $G=U(1)$, however, this is not so, and $H^2(BU(1),\Uo)=U(1)=\mathbb{R}/\mathbb{Z}$.
Moreover, it was argued in \cite{CHR} that the filling factor $\nu$ mod 1 specifies the element in $H^2(BU(1),\Uo)$.

In this sense, the formulation given in \cite{CHR} is more general and unifies two lines of generalizations of the LSM theorem, namely to  U(1) and to finite groups and compact connected semisimple groups.
It would be interesting to look for a mathematically rigorous formulation which covers both these cases simultaneously.

\section{A theorem for reflection invariant models with U(1) symmetry}
\label{app-U(1)}
We here describe and prove a theorem corresponding to Corollary~\ref{C:ref} or Theorem~\ref{T:ref} obtained by the original method of U(1) twist devised by Lieb, Schultz, and Mattis \cite[Appendix B]{LSM}.
Let us consider a standard quantum spin chain on $\bbZ$ as in section~\ref{s:intro}.
We let the spin quantum number at site $x$ be $\bar{S}_x\in\{\frac{1}{2},1,\frac{3}{2},\ldots,\bar{S}_{\rm max}\}$, and assume the symmetry $\bar{S}_x=\bar{S}_{-x}$.
The spin chain is described by the formal Hamiltonian $H=\sum_{x\in\bbZ}h_x$, where the local Hamiltonian $h_x$ acts nontrivially only on sites $y$ such that $|y-x|\le R$, and satisfies $\snorm{h_x}\le B$.
We further assume that $h_x$ is U(1) invariant, i.e.,
\eq
\exp\bigl[i\theta\sum_{y:|y-x|\le R}S^{(3)}_x\bigr]\,h_x\,\exp\bigl[-i\theta\sum_{y:|y-x|\le R}S^{(3)}_x\bigr]=h_x,
\lb{hxU(1)}
\en
for any $x\in\bbZ$ and $\theta\in[0,2\pi)$.
We also define a linear $*$-automorphism $\tcR$ by $\tcR(S^{(1)}_x)=S^{(1)}_{-x}$ and $\tcR(S^{(\nu)}_x)=-S^{(\nu)}_{-x}$ for $\nu=2,3$, for all $x\in\bbZ$.
Note that $\tcR$ describes a $\bbZ_2$ transformation which consists of reflection about the origin and the $\pi$-rotation about the 1-axis.
We make an essential assumption that $\tcR(h_x)=h_{-x}$ for all $x\in\bbZ$.
The model has a $\Uo\rtimes\bbZ_2$ symmetry.

Note that the $\bbZ_2$ symmetry in this case is not an on-site symmetry, but is a global symmetry described by $\tcR$.
This means that the present class of models is not covered by Corollary~\ref{C:ref} or Theorem~\ref{T:ref}.\footnote{%
As a special case, one can consider spin chains which have both on-site $\Uo\rtimes\bbZ_2$ symmetry and reflection symmetry.
Corollary~\ref{C:ref} certainly applies to such models because $\ZZ\subset\Uo\rtimes\bbZ_2$.
}

The main result of the present Appendix is the following.
\begin{T}\label{T:refU1}
Suppose that the above quantum spin chain has a unique ground state $\omega$.
When $\bar{S}_0$ is a half-odd integer, there is no gap above the ground state.
\end{T}

\noindent\underline{Proof}:
We follow Affleck and Lieb \cite{AL} and define the local twist operator
\eq
U_\ell\coloneqq\exp\Bigl[-i\sum_{x=-\ell}^\ell\frac{x+\ell}{\ell}\pi S^{(3)}_x\Bigr],
\lb{localtwist}
\en
for $\ell=1,2,\ldots$.
From the standard argument based on the U(1) invariance of $h_x$, we find for any $\ell$ that
\eq
\bigl|\omega(U^\dagger_\ell H_LU_\ell)-\omega(H_L)\bigr|\le\frac{C}{\ell},
\lb{UHU}
\en
with a constant $C$, where $H_L=\sum_{x=-L}^Lh_x$ with any $L\ge\ell+R$.
See, e.g., \cite{TasakiLSM,TasakiBook}.
We also note that
\eq
\tcR(U_\ell)=\exp\Bigl[i\sum_{x=-\ell}^\ell\frac{-x+\ell}{\ell}\pi S^{(3)}_x\Bigr]
=\exp\Bigl[2\pi i\sum_{x=-\ell}^\ell S^{(3)}_x\Bigr]\,U_\ell
=e^{2\pi i S^{(3)}_0}\,U_\ell,
\lb{localtwist2}
\en
which implies $\tcR(U_\ell)=-U_\ell$ when $\bar{S}_0$ is half-odd integral.\footnote{The same observation was made in \cite{MarstonTsai}.}
Since $\omega$ is $\tcR$-invariant, this means that
\eq
\omega(U_\ell)=0.
\lb{oU=0}
\en
It is standard that \rlb{oU=0} and \rlb{UHU} imply that there is no gap above the ground state.~\qedm

\bibliographystyle{ytamsalpha}
\baselineskip=.95\baselineskip
\bibliography{ref}

\end{document}